\documentclass[a4paper]{article}

\usepackage{INTERSPEECH2021}
\usepackage{amsmath,graphicx,amssymb}
\usepackage{enumitem}
\usepackage{mathtools}

\usepackage{balance,cite}

\usepackage{tabularx, etoolbox, booktabs}
\usepackage{multirow} 
\usepackage{subcaption}
\usepackage{color}
\usepackage{algorithm}
\usepackage{algpseudocode}

\newcommand{\basicbskip}{\baselineskip 9.40pt}

\ninept

\title{Advances in integration of end-to-end neural and clustering-based diarization for real conversational speech \vspace{-1mm}}

\name{Keisuke Kinoshita, Marc Delcroix, Naohiro Tawara\vspace{-1mm}}
\address{NTT Corporation, Japan}
\email{keisuke.kinoshita@ieee.org}

\begin{document}
\basicbskip

\maketitle
\begin{abstract}
Recently, we proposed a novel speaker diarization method called End-to-End-Neural-Diarization-vector clustering (EEND-vector clustering) 
that integrates clustering-based and end-to-end neural network-based diarization approaches into one framework.
The proposed method combines advantages of both frameworks, i.e. high diarization performance and handling of overlapped speech based on EEND,
and robust handling of long recordings with an arbitrary number of speakers based on clustering-based approaches. However, the method was only evaluated so far on simulated 2-speaker meeting-like data.
This paper is to (1) report recent advances we made to this framework, 
including newly introduced robust constrained clustering algorithms,
and (2) experimentally show that the method can now outperform competitive diarization methods 
such as Encoder-Decoder Attractor (EDA)-EEND,
on CALLHOME data which comprises real conversational speech data including overlapped speech and an arbitrary number of speakers.
By further analyzing the experimental results, 
this paper also discusses pros and cons of the proposed method and reveals potential for further improvement.
A set of the code to reproduce the results is available at \cite{eend-vc_code}.
\end{abstract}

\def\nspk{K}
\def\nspki{k}
\def\nout{N}
\def\nouti{n}
\def\nutt{U}
\def\nutti{u}
\def\signal{\vect{s}}
\def\estimate{\hat{\signal}}
\def\coloringset{\mathcal{C}}
\def\graph{G}
\def\shift{l}
\def\meetingsignal{\vect{y}}
\def\historycontext{N_\text{h}}
\def\futurecontext{N_\text{f}}
\def\currentcontext{N_\text{c}}

\section{Introduction}
\label{sec:intro}
Automatic meeting/conversation analysis is one of the essential technologies required 
for realizing futuristic speech applications such as communication agents that can follow, respond to, and facilitate our conversation. 
As an important central task for the meeting analysis, speaker diarization has been extensively studied \cite{Diarization_review, DIHARD_data, AMI_data}.

Currently, there are mainly two major approaches to the diarization problem, that is,
clustering-based approaches \cite{x-vector,Diarization_review,DIHARD_JHU,DIHARD_BUT} such as x-vector clustering,
and End-to-End Neural Diarization (EEND) approaches \cite{Fujita_IS2019,Fujita_ASRU2019,Horiguchi2020_EDA_EEND}.
The clustering-based approaches first segment a recording into short homogeneous blocks 
and compute a speaker embedding for each block 
assuming that only one speaker is active in each block. 
Then, speaker embedding vectors are clustered to regroup segments belonging to the same speakers and obtain diarization results \cite{DIHARD_JHU,DIHARD_BUT,Zhang_ICASSP19, Li_2020_RelationNet}.
On the other hand, EEND is relatively simple. It receives standard frame-level spectral features and directly outputs a frame-level speaker activity
for each speaker.
In recent diarization challenges such as DIHARD-III \cite{DIHARD_3}, 
it is revealed that these two approaches are complementary to each other as it will be discussed below, 
and thus many institutes achieved reliable diarization for real conversational data by performing system combination of these approaches, e.g., \cite{Hitachi_JHU_DIHARD3}.

To accomplish reliable diarization for any real conversational speech, 
the following essential problems have to be addressed:
\begin{itemize}
   \item[(1)] overlapped speech (i.e., segments where more than one person is speaking)
   \item[(2)] long-form audio (e.g., duration of more than 10 min.), 
   \item[(3)] an arbitrary number of speakers.
\end{itemize}
In each of these aspects, the aforementioned two approaches have different properties 
complementary to each other.
The clustering-based approaches have been studied for a decade 
and have been shown to work well with long-form audio containing an arbitrary number of speakers \cite{Diarization_review}.
However, by nature of the assumption made in the extraction of speaker embeddings, i.e., single-speaker block assumption,
it cannot handle overlapped speech.
On the other hand,
EEND was first developed to address the overlapped speech problem \cite{Fujita_IS2019,Fujita_ASRU2019}.
Then, recently, it was extended to handle meetings containing an arbitrary number of speakers 
by introducing speaker counting functionality based on an Encoder-Decoder Attractor architecture (EDA) \cite{Horiguchi2020_EDA_EEND}.
However, it was experimentally shown to still have difficulty in dealing with a meeting containing a realistically large number of speakers, 
such as more than 3 speakers \cite{Horiguchi2020_EDA_EEND}.
In addition, it was shown that 
it is difficult to directly apply the EEND systems to long-form audio 
(e.g., recordings longer than 10 minutes) \cite{EEND-vector-clustering_ICASSP2021}.
Since the original EEND system was designed to operate in a batch processing mode, 
it inevitably requires a very large computer memory when performing inference with long recordings.
Besides, aside from the memory issue, the neural networks (NNs) in EEND have difficulty generalizing to unseen very long sequential data.
Block-wise independent processing is also difficult because it poses an inter-block label permutation problem, i.e., an ambiguity of the speaker label assignments between blocks.

Focusing on these different pros and cons of the clustering and EEND approaches,
we proposed a 
simple but effective hybrid diarization approach \cite{EEND-vector-clustering_ICASSP2021},
called EEND-vector clustering, by combining the best of the clustering-based diarization and EEND.
The framework allows us to process long-form audio containing overlapped speech and an arbitrary number of speakers.
It first split the input long recording into fixed-length blocks.
Then it applies a modified version of EEND to each block
to obtain the diarization results for a fixed number of speakers 
as well as global speaker embedding vectors for each of the speakers. 
This assumes that for each short block the maximum number of speakers 
will be equal to or less than the number of the output of EEND.
Finally, to solve the inter-block label permutation problem,
speaker clustering is performed across blocks by using a constrained clustering algorithm.
In \cite{EEND-vector-clustering_ICASSP2021},
the EEND-vector clustering framework was shown to significantly outperform the conventional EEND \cite{Fujita_ASRU2019} and x-vector clustering
when processing {\it simulated} long-recordings of 2 speakers containing various overlap conditions, 
noise and reverberation,
and thus was proven to be more advantageous in addressing the aforementioned problems (1) and (2).

However, it was not clear from our past studies 
whether the EEND-vector clustering could be generalized 
to {\it real} conversational speech data containing an arbitrary number of speakers (aforementioned problems (1) and (3)).
To this end, this paper focuses on 
(i) application of the EEND-vector clustering 
to the widely used CALLHOME dataset \cite{CALLHOME}, which consists of the real conversational speech of 2 to 6 speakers,
and (ii) its evaluation in comparison with 
current state-of-the-art systems such as EDA-EEND \cite{Horiguchi2020_EDA_EEND},
x-vector clustering \cite{DIHARD_BUT}, 
and Region-Proposal Network based Speaker Diarization (RPNSD) \cite{Region_proposal_NN_diarization}
that can handle CALLHOME data including overlapped speech.
We also (iii) introduce practical techniques to increase robustness against real data  
such as more robust constrained clustering methods and silent speaker detection.

In the remainder of the paper, we first review the proposed EEND-vector clustering approach (in Sec.~2),
and then introduce the practical techniques required to deal with real meeting data (in Sec.~3).  
Finally, with experiments,
we show that the EEND-vector clustering can outperform the other state-of-the-art approaches by a large margin, 
especially when the number of speakers is large.

\begin{figure}[t]
 \begin{center}
  \includegraphics[width=80mm]{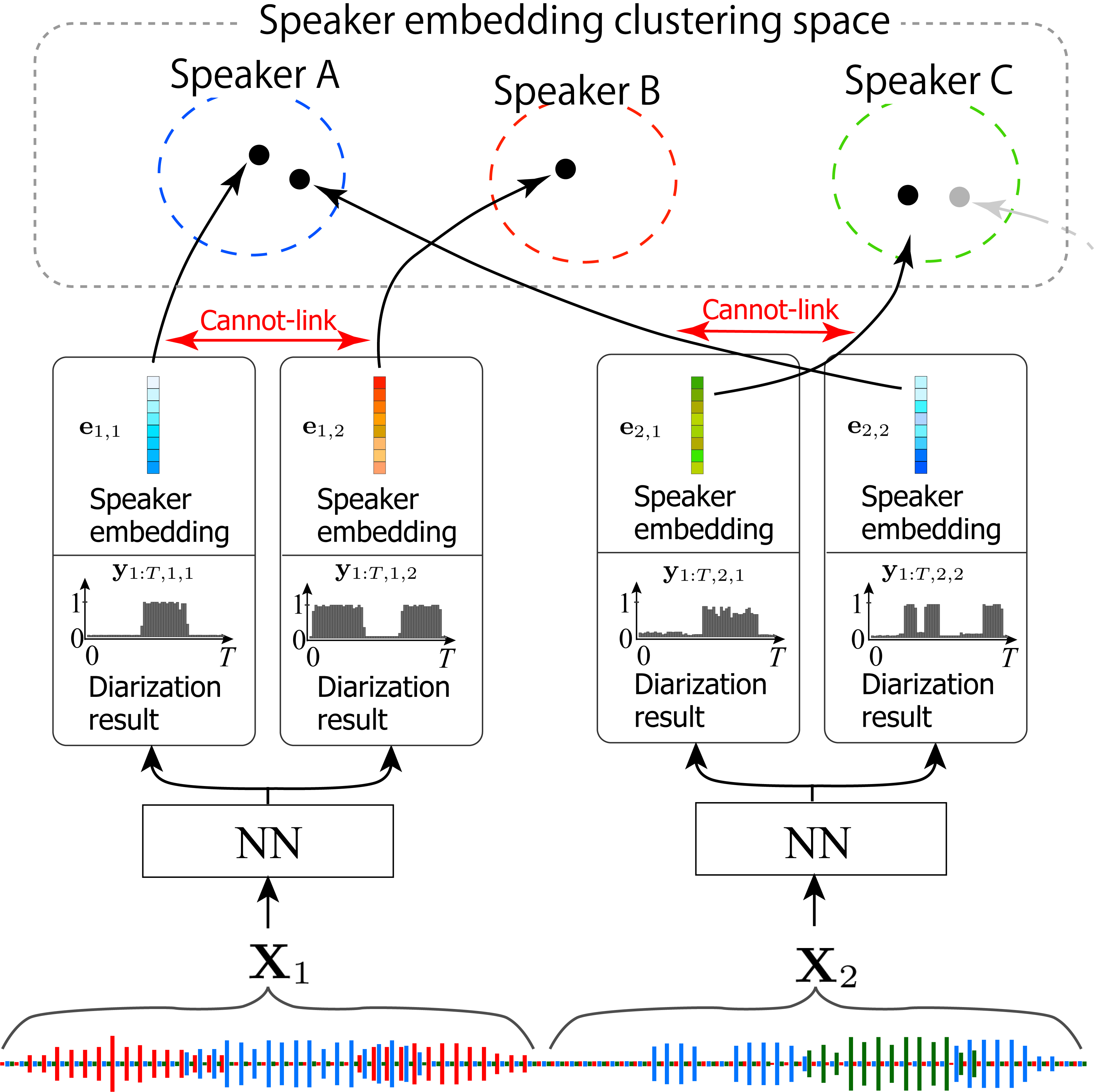}
     \end{center}
   \vspace{-3mm}
   \caption{Schematic diagram of the proposed diarization framework. The input contains 3 speakers in total (red, green, and blue speakers shown in the waveform at the bottom), but only at most 2 speakers are actively speaking in each block.}
 \label{fig:overview}
\end{figure}

\section{EEND-vector clustering}
\label{sec:proposed_method}

\subsection{Overall framework}
\label{sec:framework}
Figure~\ref{fig:overview} shows a schematic diagram of the proposed EEND-vector clustering framework, originally proposed in \cite{EEND-vector-clustering_ICASSP2021}.

It first segments the input recording into blocks 
and calculates a sequence of the input frame features within each block, 
as $\mathbf{X}_i = (\mathbf{x}_{t,i} \mid t=1,\cdots,T)$ where $i$,$t$ and $T$ are the block index, 
the frame index in the block and the block size, respectively.
$\mathbf{x}_{t,i} \in \mathbb{R}^{K}$ is the $K$-dimensional input frame feature at the time frame $t$.
In the example shown in Fig~\ref{fig:overview}, the input recording consists of 2 blocks and contains 3 speakers in total. 
In the following explanation, we assume that we can reasonably fix the maximum number of active speakers 
in a block, $S_{\textrm{Local}}$, to 2, for the sake of simplicity
\footnote{In experiments, we use $S_{\textrm{Local}}=3$.}
.

Based on the assumption/hyper-parameter $S_{\textrm{Local}}=2$,
the neural network $\mathrm{NN}$ always estimates diarization results 
and associated speaker embeddings for 2 speakers in each block.
If a speaker is absent (i.e., there is only one active speaker in that block), 
the network simply estimates the diarization results of all zeros for that silent speaker.
The diarization results are estimated independently in each block 
\footnote{For the details of NN and its training procedure, please refer to \cite{EEND-vector-clustering_ICASSP2021}.}.
Since it is {\it not} always guaranteed that the diarization results of a certain speaker are estimated at the same output node,
we may have the inter-block label permutation problem in the diarization outputs.
We can solve this permutation problem and estimate the correct association of the diarization results among blocks, 
by clustering the speaker embeddings given the total number of speakers in the input recording, $S_{\textrm{Global}}$, (3 in this case).
Note that the speaker embedding extraction process is optimized 
such that the vectors of the same speaker stay close to each other,
while the vectors of different speakers lie far away from each other.
Based on the clustering results,
we can stitch together the diarization results of the same speaker across blocks to obtain the final diarization output.
Note that, while the proposed framework estimates the diarization results for the fixed number of speakers $S_{\textrm{Local}}$ in a block, 
it can handle a meeting with an arbitrary number of speakers.

To tightly couple the embedding estimation and clustering process,
it is beneficial to convey to the clustering algorithm the useful information 
that NN always estimates speech activities of two {\it different} speakers in each block.
To this end, we suggest using a {\it constrained/semi-supervised} clustering algorithm which typically allows us to set a cannot-link constraint 
between a given pair of embeddings to prevent the pair from being assigned to the same speaker cluster.

\subsection{Formulation of NN to jointly perform diarization \\ 
and speaker embedding estimation}
$\mathrm{NN}$ in Fig.~\ref{fig:overview} can be formulated as follows. 
Let us denote the ground-truth diarization label sequence as 
$\mathbf{Y}_i = \{\mathbf{y}_{t,i} \mid t=1,\cdots,T \} \in \mathbb{R}^{S_{\textrm{Local}} \times T}$ that corresponds to $\mathbf{X}_i$.
Here, the diarization label $\mathbf{y}_{t,i} = [y_{t,i,s} \in \{0,1\} \mid s=1, \cdots, S_{\textrm{Local}}]$ represents a joint activity for $S_{\textrm{Local}}$ speakers.
For example, $y_{t,i,s} = y_{t,i,s'} = 1 (s \ne s')$ indicates both speakers $s$ and $s'$ spoke at the time frame $t$ in the block $i$. 
Similarly, let us denote the ground-truth speaker embedding set as 
$\mathbf{E}_i = (\mathbf{e}_{i,s} \mid s=1,\dots,S_{\textrm{Local}}) \in \mathbb{R}^{S_{\textrm{Local}} \times C}$ 
that corresponds to $\mathbf{X}_i$.
$\mathbf{e}_{i,s}$ is $C$-dimensional speaker embedding vector for the $s$-th speaker.

Then, the joint estimation of diarization results and speaker embeddings by NN in Fig.~\ref{fig:overview} 
is formulated as:
\begin{align}
    \hat{\mathbf{Y}}_i,\hat{\mathbf{E}}_i &= \mathrm{NN}( \mathbf{X}_i ), \nonumber 
\end{align}
$\mathrm{NN}$ can be trained with a multitask loss function composed of diarization loss, i.e., binary cross entropy loss,
and speaker embedding loss that encourages the embeddings to have small intra-speaker and large inter-speaker euclidean distances,
as proposed in \cite{EEND-vector-clustering_ICASSP2021}.

\section{Handling real conversational data}
\label{sec:practical}
This section summarizes techniques that we incorporated into the EEND-vector clustering 
to cope with real conversational data.
While we should be able to improve the NN architecture and training procedure 
as in much other literature \cite{Fujita_ASRU2019,Horiguchi2020_EDA_EEND} to push up the final performance, 
we found that improving only the constrained clustering part, which is a unique part of EEND-vector clustering,
can already make a big difference in the final performance and help achieve state-of-the-art performance.
The following subsections detail modifications we newly introduce to the original framework proposed in \cite{EEND-vector-clustering_ICASSP2021}.

\subsection{Improving constrained clustering}
In \cite{EEND-vector-clustering_ICASSP2021}, we employed a constrained clustering algorithm called COP-Kmeans \cite{COP-kmeans}
which is an extension of k-means clustering, and showed it was effective for handling simulated 2-speaker noisy reverberant meeting data.
However, we neither compared it with other constrained clustering algorithms nor confirmed whether it is beneficial to use the cannot-link constraint. 
To thoroughly investigate the effectiveness of the constrained clustering algorithm in the proposed framework,
here we incorporate and evaluate other constrained clustering algorithms in addition to COP-Kmeans.
While the standard k-means clustering optimizing squared error generally does not well handle non-Gaussian data and/or data containing imbalanced classes, 
both of which are common for real conversational speech data,
it was shown that Agglomerative Hierarchical Clustering (AHC) do not suffer from such limitations 
and thus has been widely used within clustering-based diarization systems \cite{DIHARD_JHU,DIHARD_BUT}.
Also, Spectral Clustering (SC) is often used for clustering-based diarization systems \cite{Zhang_ICASSP19, IBM_diarization},
since it can handle non-Gaussian data.
In accordance with this development in the clustering for diarization, 
we here introduce the following constrained AHC and constrained SC into the EEND-vector clustering framework.

\subsubsection{Constrained Agglomerative Hierarchical Clustering}
AHC is a common but effective unsupervised clustering technique used 
in state-of-the-art diarization systems \cite{DIHARD_JHU,DIHARD_BUT}.
Standard AHC requires pairwise distances between all input samples, i.e., speaker embeddings $\hat{\mathbf{E}}_i\ (i=1,\ldots,I)$, 
resulting in a distance matrix. 
Then, based on the matrix, 
it generates a cluster hierarchy that is commonly displayed as a tree diagram called a dendrogram.
Each input sample starts in its own cluster, and pairs of clusters are merged as it moves up the hierarchy.
We stop the cluster merging process either (a) when we obtain a required number of clusters, 
or (b) when the clusters are too far apart beyond a certain threshold to be merged.
In diarization systems, the stopping criterion (a) is used to obtain diarization results 
given prior knowledge on the number of speakers in a meeting (i.e., oracle-number-of-speaker evaluation),
while the stopping criterion (b) is used to obtain diarization results 
with the estimated number of speakers in a meeting (i.e., estimated-number-of-speaker evaluation).

To incorporate the cannot-link constraint to AHC, it is proposed to directly modify the distance matrix
such that the distance between samples with the cannot-link constraint becomes larger than any other values
in the matrix \cite{constrained_AHC}. 
Within the proposed framework, the distance matrix $\bf{D}^{\mathrm{dist.}} \in \mathbb{R}^{I\cdot S_{\textrm{Local}} \times I \cdot S_{\textrm{Local}}}$ we obtain based on $\hat{\mathbf{E}}_i\ (i=1,\ldots,I)$ can be expressed as
\begin{align}
\bf{D}^{\mathrm{dist.}} &= 
\begin{bmatrix}
\bf{D}_{1,1} & \cdots &  \bf{D}_{1,I} \\
\vdots       & \ddots &  \vdots       \\
\bf{D}_{I,1} & \cdots &  \bf{D}_{I,I}  
\end{bmatrix} \label{eq:distance_matrix}
\end{align}
where each element of the above block matrix, $\bf{D}_{i,j} \in \mathbb{R}^{S_{\textrm{Local}} \times S_{\textrm{Local}}}$, 
is a symmetric distance matrix calculated 
based on a set of speaker embedding vectors obtained at $i$-th block, $\hat{\mathbf{E}}_i$,
and $j$-th block, $\hat{\mathbf{E}}_j$.
When incorporating the cannot-link constraint,
we insert certain large value $\kappa$ into the off-diagonal components of matrices $\bf{D}_{i,i}$ for all $i$ from $1$ to $I$.
After obtaining the modified distance matrix, we can use it in a standard AHC
with the aforementioned stopping criteria.

\subsubsection{Constrained Spectral Clustering}
SC is another common unsupervised clustering technique used 
in recent diarization systems \cite{Zhang_ICASSP19, IBM_diarization}.
SC is a technique with roots in graph theory, and it requires a pairwise similarity score between all input sample, 
i.e., speaker embeddings $\hat{\mathbf{E}}_i\ (i=1,\ldots,I)$, resulting in a similarity graph.
When we have prior knowledge on the number of speakers, $S_{\textrm{Global}}$, in a meeting,
we typically analyze the eigenvectors corresponding to $S_{\textrm{Global}}$ smallest eigenvalues of unnormalized graph Laplacian constructed from the similarity graph.
When we would like to estimate the number of clusters,
we can use a widely used method called eigengap heuristics \cite{Spectral_clustering_tutorial}.

There are several ways to incorporate the cannot-link constraint to SC \cite{constrained_SC},
varying from a simple way \cite{Spectral_learning_Kamvar} 
to methods that can control the contribution of the constraint in a soft manner \cite{constrained_SC}.
Here we employ a relatively straightforward method proposed in \cite{Spectral_learning_Kamvar}, 
which directly modifies the similarity graph such that the graph edge between samples with the cannot-link is forced to $0$.
This can be done by constructing the similarity matrix similar to eq.~(\ref{eq:distance_matrix}) but with similarity scores, and inserting 
$0$ into the off-diagonal components of matrices $\bf{D}_{i,i}$ for all $i$ from $1$ to $I$.

\subsection{Silent speaker detection}
Before clustering speaker embedding vectors with a constrained clustering,
it is beneficial to detect and exclude embedding vectors corresponding to silent speakers.
It is mainly because, if there are multiple silent speakers in that block, we should not set the cannot-link constraint between those embedding vectors, 
i.e., they should belong to the same silent speaker cluster.
We found that, in many cases, the diarization results for those silent speakers stay very close to $0$ (as we trained NN to do so).
Therefore, we propose to detect it by examining whether the mean of the diarization results is sufficiently small; 
A speaker embedding is judged to be from a silent speaker if $\frac{1}{T} \sum_{t=1}^T \hat{y}_{t,i,s} < \tau $,
where $\tau$ is a predetermined threshold.

\section{Experiments}
In this section, we evaluate the effectiveness of the proposed EEND-vector clustering 
in comparison with state-of-the-art conventional methods, 
based on CALLHOME dataset \cite{CALLHOME}. 
We also evaluate clustering algorithms in the proposed EEND-vector clustering framework 
to highlight the importance of the constrained clustering.
A set of the code to reproduce the following results of the EEND-vector clustering is available at \cite{eend-vc_code}.

\subsection{Data}
For the training, we used simulated mixtures created from Switchboard-2 (Phase I \& I\hspace{-.1em}I \& I\hspace{-.1em}I\hspace{-.1em}I), Switchboard Cellular (Part 1 \& 2), and the NIST Speaker Recognition Evaluation (2004 \& 2005 \& 2006 \& 2008) for speech,
and the MUSAN corpus \cite{MUSAN} for noise with simulated room impulse responses used in \cite{Ko_2017},
by following the data generation procedure in \cite{Fujita_IS2019}.
We created 3-speaker meeting-like dataset based on the algorithm proposed in \cite{Fujita_IS2019} with $\beta=10$
. 
\footnote{Each mixture contains dozens of utterances per speaker with reasonable silence intervals between utterances of the same speaker's. 
$\beta=10$ means that the average duration of the silence interval is 10~s.
}

For evaluation and adaptation, we used the telephone conversation dataset CALLHOME (CH) \cite{CALLHOME}, 
i.e., NIST SRE 2000 (LDC2001S97, Disk-8), which has been the most widely used dataset for speaker diarization studies.
The CALLHOME dataset contains 500 sessions of multilingual telephonic speech. Each session has 2 to 6 speakers 
while there are two dominant speakers in each conversation.
For evaluation and adaptation purpose, 
we split the CALLHOME dataset into two subsets according to \cite{Horiguchi2020_EDA_EEND},
and performed adaptation on a subset and evaluation of the proposed method on the other subset.

\subsection{Conventional methods to be compared with}
The proposed method was compared with state-of-the-art methods, 
namely, x-vector clustering with PLDA scoring and variational Bayesian resegmentation \cite{x-vector, Diarization_review, DIHARD_BUT}, 
EDA-EEND \cite{Horiguchi2020_EDA_EEND} and Region-Proposal Network based Speaker Diarization (RPNSD) \cite{Region_proposal_NN_diarization}.
We did not reproduce their results, but simply borrow the results of x-vector clustering and EDA-EEND from \cite{Horiguchi2020_EDA_EEND},
and that of RPNSD from \cite{Region_proposal_NN_diarization}.

\subsection{Settings of the proposed EEND-vector clustering}
\subsubsection{NN training and hyper-parameters}
In this experiment, the assumed maximum number of speakers in each block $S_{\textrm{Local}}$ was set at $3$ in the proposed method,
i.e., $\mathrm{NN}$ always estimates diarization results and embedding vectors for 3 speakers including silent speaker(s).
The block size $T$ was varied from 50~s to 15~s to see how the block size will have an impact on the diarization performance.
To speed up the experiments, we first trained a model with $S_{\textrm{Local}}=3$ and $T=15~\mathrm{s}$ on the training data for 100 epochs,
and adapted the pre-trained model to different block sizes $T$s to obtain a model appropriate for each $T$.

For the neural network architecture and training protocol,
we basically followed \cite{Horiguchi2020_EDA_EEND}, and followed \cite{EEND-vector-clustering_ICASSP2021} for speaker embedding estimation
and multi-task training setting.
We used self-attention-based six-layer stacked Transformer encoders with eight attention heads as a backbone of our method.
The input for the network was the same as \cite{Horiguchi2020_EDA_EEND}, i.e., 345-dimensional log-scaled Mel-filterbank-based features.
The threshold $\tau$ to detect the silent speaker was set at 0.05.

\subsubsection{Clustering algorithms in the proposed method}
We evaluate oracle clustering, the following 3 unconstrained and 3 constrained clustering algorithms 
in the framework of the EEND-vector clustering, namely,
k-means, AHC, SC, COP-Kmeans, constrained AHC, and constrained SC.

The oracle clustering corresponds to permutation that can yield a diarization result closest to the true one, 
based on a diarization result estimated by the network.

Using unconstrained clustering algorithms such as k-means, AHC, and SC,
there may be some cases where speaker embeddings from a certain block are clustered into the same cluster.
In such case, with an assumption that a certain speaker's speech activity was erroneously split into more than one output,
we heuristically merge the diarization results corresponding to those speaker embeddings,
by taking maximum across these diarization results at each time frame $t$.

For the constrained AHC, we set $\kappa$ at $10000$.
We performed the AHC clustering such that it minimizes the average of the distances between all observations of pairs of clusters.
The distance threshold above which, clusters will not be merged, was set at 1.

\begin{table}[t]
    \centering
    \caption{DER (\%) of EEND-vector clustering with various clustering algorithms for CALLHOME dataset.}
    \vspace{-3mm}
    \label{tbl:clustering}
    \resizebox{\linewidth}{!}{
    \begin{tabular}{@{}lcccccc|c@{}}
        \toprule
        & & \multicolumn{6}{c}{\# of speakers in a session}\\\cmidrule(l){3-8}
        Method                      & speaker counting &  2 & 3 & 4 & 5 & 6 & All\\\midrule \midrule
        Oracle clustering                                   & oracle         &    7.51    &    9.66    &    12.34    &    12.93    &    22.01    &    10.21 \\
        \midrule
        Kmeans                                              & oracle         &    8.54    &    21.22    &    24.79    &    34.16    &    31.97    &    18.68 \\
        COP-Kmeans \cite{EEND-vector-clustering_ICASSP2021} & oracle         &    9.25    &    19.55    &    24.49    &    35.71    &    41.03    &    18.74 \\
        SC                                                  & oracle         &    7.95    &    13.07    &    19.45    &    23.04    &    35.54    &    14.12 \\
        Constrained SC                                      & oracle         &    7.94    &    12.17    &    19.56    &    20.12    &    35.95    &    13.73 \\
        AHC                                                 & oracle         &    7.96    &    11.90    &    15.16    &    25.01    &    28.36    &    12.59 \\
        Constrained AHC                                     & oracle         &    8.08    &    11.27    &    15.01    &    23.14    &    26.56    &    \bf{12.22} \\
        \midrule
        SC                                                  & estimated      &    8.13    &    16.24    &    23.13    &    37.62    &    35.67    &    16.75 \\
        Constrained SC                                      & estimated      &    8.71    &    16.25    &    23.51    &    37.62    &    43.10    &    17.38 \\
        AHC                                                 & estimated      &    9.54    &    12.49    &    21.49    &    24.41    &    32.18    &    14.78 \\
        Constrained AHC                                     & estimated      &    7.96    &    11.93    &    16.38    &    21.21    &    23.10    &    \bf{12.49} \\
      \bottomrule
    \end{tabular}
    }
\end{table}

\subsection{Results}
\subsubsection{Effect of clustering algorithms}
Table~\ref{tbl:clustering} shows diarization error rate (DER) of the proposed EEND-vector clustering with best-performing condition ($T=30s$),
with different clustering methods performed with the oracle number of speakers and the estimated number of speakers.
First, we can see that the constrained AHC performs the best and stays closest to the oracle clustering.
In comparison with its unconstrained counterpart, it performs significantly better for difficult cases such as the number of speakers of 6.
Hereafter, in this paper, we use constrained AHC as the clustering algorithm in the proposed method.

Except for AHC, it is not perfectly clear whether the incorporation of the cannot-link constraint 
is advantageous or not. The obtained results are very much affected by issues such as the imbalanced class problem,
poor accuracy in speaker counting, which makes it difficult to make a definitive conclusion. 



\begin{table}[t]
    \centering
    \caption{DERs (\%) on CALLHOME diarization results with oracle number of speakers. $T$ values in parentheses are the block sizes in EEND-vector clustering.}
    \label{tbl:overall_result_oracle}
    \vspace{-3mm}
    \resizebox{\linewidth}{!}{
    \begin{tabular}{@{}lccccc|c@{}}
        \toprule
        &\multicolumn{6}{c}{\# of speakers in a session}\\\cmidrule(l){2-7}
        Method &  2 & 3 & 4 & 5 & 6 & All\\\midrule
        x-vector clustering \cite{Horiguchi2020_EDA_EEND}                   &8.93       &19.01       &24.48       & 32.14      & 34.95      & 18.98\\
        RPNSD \cite{Region_proposal_NN_diarization}                        &N/A        &N/A         &N/A         &N/A         &N/A         & 17.06 \\
        EDA-EEND \cite{Horiguchi2020_EDA_EEND} & 8.35 & 13.20 & 21.71      & 33.00      & 41.07      & 15.43\\
        \midrule
        EEND-vector clust. ($T=50s$)                    &    \bf{7.63}    &    13.14    &    \bf{13.71}    &    \bf{22.14}    &    28.82    &    12.53 \\
        EEND-vector clust. ($T=30s$)                    &    8.08    &    \bf{11.27}    &    15.01    &    23.14    &    26.56    &    \bf{12.22} \\
        EEND-vector clust. ($T=20s$)                    &    7.97    &    13.56    &    15.28    &    28.52    &    \bf{22.62}    &    13.15 \\
        EEND-vector clust. ($T=15s$)                    &    8.57    &    14.33    &    17.12    &    30.02    &    23.17    &    14.07 \\
      \bottomrule
    \end{tabular}
    }
\end{table}

\begin{table}[t]
    \centering
    \caption{DERs (\%) on CALLHOME diarization results with estimated number of speakers. $T$ values in parentheses are the block sizes in EEND-vector clustering.}
    \vspace{-3mm}
    \label{tbl:overall_result_estimated}
    \resizebox{\linewidth}{!}{
    \begin{tabular}{@{}lccccc|c@{}}
        \toprule
        &\multicolumn{6}{c}{\# of speakers in a session}\\\cmidrule(l){2-7}
        Method                 &  2 & 3 & 4 & 5 & 6 & All\\\midrule
        x-vector clustering \cite{Horiguchi2020_EDA_EEND}                       &15.45      &18.01          & 22.68      & 31.40       &34.27       & 19.43\\
        EDA-EEND \cite{Horiguchi2020_EDA_EEND}      & 8.50      & 13.24    & 21.46      & 33.16       & 40.29      & 15.29\\
        \midrule
        EEND-vector clust. ($T=50s$)                    &    \bf{7.18}    &    12.50    &    16.91    &    28.04    &    26.76    &    12.98 \\
        EEND-vector clust. ($T=30s$)                    &    7.96    &    \bf{11.93}    &    16.38    &    \bf{21.21}    &    23.10    &    \bf{12.49} \\
        EEND-vector clust. ($T=20s$)                    &    8.13    &    12.83    &    16.75    &    31.90    &    \bf{23.08}    &    13.40 \\
        EEND-vector clust. ($T=15s$)                    &    9.77    &    16.27    &    \bf{16.12}    &    27.21    &    23.16    &    14.84 \\
      \bottomrule
    \end{tabular}
    }
\end{table}

\subsubsection{Comparison with state-of-the-art methods}
Tables~\ref{tbl:overall_result_oracle} and \ref{tbl:overall_result_estimated} shows diarization error rate (DER) of the proposed EEND-vector clustering
in comparison with state-of-the-art methods, performed with the oracle number of speakers (Table~\ref{tbl:overall_result_oracle}) 
and with the estimated number of speakers (Table~\ref{tbl:overall_result_estimated}).
They first reveal that, comparing the best performing configuration of the proposed method ($T=30s$) with the other state-of-the-art methods,
EEND-vector-clustering largely outperforms the others.
It works significantly better than EDA-EEND especially when the number of speakers is large.
The proposed method works generally better than x-vector-clustering most probably because it handles overlapped speech well
and can estimate speaker embeddings from longer segments.
It also outperforms RPNSD that combines NN-based diarization with clustering in a different manner.

However, if we decrease the size of the processing block $T$, the diarization performance tends to degrade. 
This issue should be resolved in the future, because, although 
for telephone speech data such as CALLHOME, speakers tend to speak for a long time and there are relatively few different speakers appearing in relatively long segments, it may not be always the case, especially when dealing with more causal real-life conversations.
To deal with such data, we need to use a shorter block size to keep the maximum number of speakers in a block as small as $S_{\textrm{Local}}=3$. 
It is not shown here because of space limitation, 
but, by examining the DERs more carefully, we found that, as we decrease the block size, 
missed speech and false alarm do not significantly increase but speaker confusion does.
It suggests that if we could improve accuracy of the speaker embeddings or the clustering algorithm,
we could bring about significant improvement even when the processing block is short, which will be part of our future works.

\section{Conclusions}
This paper evaluated the EEND-vector clustering framework based on real conversational speech dataset CALLHOME,
and showed that it outperforms significantly the conventional state-of-the-art methods.
We also experimentally showed the importance of constrained clustering in our framework.

\pagebreak
\balance
\bibliographystyle{IEEEtran}

\bibliography{mybib}


\end{document}